\documentclass[aps,prd,11pt,showpacs,showkeys,preprintnumbers,superscriptaddress,nobibnotes,floatfix,longbibliography,nofootinbib]{revtex4-1}

\pdfoutput=1
\usepackage{mathrsfs}
\usepackage{hyperref}
\usepackage{graphicx}
\usepackage{amsfonts,amsmath,amssymb,bm}
\usepackage{array}
\usepackage{color}
\usepackage{enumitem}
\usepackage[explicit]{titlesec}
\usepackage[utf8]{inputenc}
\usepackage{graphicx}
\usepackage{multirow}
\usepackage{lineno}
\usepackage{ulem}
\usepackage[caption=false]{subfig}

\begin{document}

\title{Environmental radon control in the 700-m underground laboratory at JUNO}

\author{Chenyang Cui}
\affiliation{Institute of High Energy Physics, Chinese Academy of Sciences, Beijing, China}
\affiliation{University of Chinese Academy of Sciences, Beijing, China}
\author{Jie Zhao}
\email{zhaojie@ihep.ac.cn}
\affiliation{Institute of High Energy Physics, Chinese Academy of Sciences, Beijing, China}
\author{Gaosong Li}
\affiliation{Institute of High Energy Physics, Chinese Academy of Sciences, Beijing, China}
\author{Yongpeng Zhang}
\affiliation{Institute of High Energy Physics, Chinese Academy of Sciences, Beijing, China}
\author{Cong Guo}
\affiliation{Institute of High Energy Physics, Chinese Academy of Sciences, Beijing, China}
\author{Zhenning Qu}
\affiliation{Institute of High Energy Physics, Chinese Academy of Sciences, Beijing, China}
\affiliation{University of Chinese Academy of Sciences, Beijing, China}
\author{Yifang Wang}
\affiliation{Institute of High Energy Physics, Chinese Academy of Sciences, Beijing, China}
\author{Xiaonan Li}
\affiliation{Institute of High Energy Physics, Chinese Academy of Sciences, Beijing, China}
\author{Liangjian Wen}
\affiliation{Institute of High Energy Physics, Chinese Academy of Sciences, Beijing, China}
\author{Miao He}
\affiliation{Institute of High Energy Physics, Chinese Academy of Sciences, Beijing, China}
\author{Monica Sisti}
\affiliation{INFN Milano Bicocca and Universit\`{a} di Milano-Bicocca, Milano, Italy}
\date{\today}

\begin{abstract}
The Jiangmen Underground Neutrino Observatory is building the world's largest liquid scintillator detector with a 20~kt target mass and about 700~m overburden. The total underground space of civil construction is about 300,000~m$^3$ with the main hall volume of about 120,000~m$^3$, which is the biggest laboratory in the world. Radon concentration in the underground air is quite important for not only human beings' health but also the background of experiments with rare decay detection, such as neutrino and dark matter experiments. The radon concentration is the main hall is required to be around 100~Bq/m$^3$. Optimization of the ventilation with fresh air is effective to control the radon underground. To find the radon sources in the underground laboratory, we made a benchmark experiment in the refuge room near the main hall. The result shows that the radon emanating from underground water is one of the main radon sources in the underground air. The total underground ventilation rate is about 160,000~m$^3$/h fresh air with about 30~Bq/m$^3$ $^{222}$Rn from the bottom of the vertical tunnel after optimization, and 55,000~m$^3$/h is used for the ventilation in the main hall. Finally, the radon concentration inside the main hall decreased from 1600~Bq/m$^3$ to around 100~Bq/m$^3$. The suggested strategies for controlling radon concentration in the underground air are described in this paper.
\end{abstract}

\maketitle

\tableofcontents
\clearpage


\section{Introduction}
\label{sec1}
Radon ($^{222}$Rn) is a well-known radioactive noble gas with a half-life of 3.8 days, produced by the alpha decay of $^{226}$Ra in the natural uranium chain. Outdoor radon concentration is typically at the level of 10~Bq/m$^3$~\cite{unscear2000}, while it can reach several thousand Bq/m$^3$ in underground locations with no special ventilation conditions. Radon in the atmosphere is the most significant contributor to human exposure from natural sources, due to the inhalation and subsequent deposition of its short-lived decay products along the walls of the respiratory tract.
Besides that, radon and its daughters are one of the most important radioactive backgrounds for neutrino and dark matter experiments expecting very low signal rate, such as JUNO~\cite{JUNO:2021vlw}, Borexino~\cite{Borexino:2019wln}, SNO+~\cite{SNO:2022qvw}, nEXO~\cite{nEXO:2021ujk}, PandaX~\cite{PandaX:2023tfq}, and DEAP-3600~\cite{DEAP:2019yzn} -- to cite only a selection of projects using liquid detector media. In order to reduce the background component due to cosmic radiation, this type of experiments are usually located in underground laboratories, where the radon concentration may represent a serious issue.

There are many underground laboratories in the world dedicated to rare event searches, among which SNOLAB in Canada and the Gran Sasso Underground Laboratory (LNGS) in Italy have relatively larger volumes compared to others~\cite{Baudis:2022pzb}. Both of them have a well designed ventilation system to achieve a low radon environment. 
The SNOLAB underground facility is located in the working Creighton nickel mine at a depth of 6000~m.w.e. (meters of water equivalent) near Sudbury, Ontario. Its total volume is equal to 470,000~m$^3$, and the underground radon concentration was reduced to 130~Bq/m$^3$ with a ventilation rate of 180,000~m$^3$/h, allowing 10 times air changes per hour in the smaller laboratory areas and 5 times air changes per hour in the three main detector cavities~\cite{SNOLAB}. LNGS is located under the Gran Sasso mountain, in the center of Italy (not far from Rome) with a rock coverage of 3800~m.w.e. Its total volume is equal to 180,000~m$^3$, and the ventilation rate reached 40,000~m$^3$/h. The radon concentration in the underground air was lowered to 20~Bq/m$^3$ in Halls A and B as of 1995, by implementing enhanced ventilation measures; however, in Hall C and the connecting tunnels, which had poorer ventilation, the radon concentration ranged from 250 to 400 Bq/m$^3$~\cite{LNGS}. In subsequent years, efforts were made to improve ventilation also in Hall C, leading to a radon concentration that is now stably around 50~Bq/m$^3$~\cite{Bucci}.

The Jiangmen Underground Neutrino Observatory (JUNO) will be the largest liquid scintillator experiment devoted to neutrino physics. The detector is made of 20~kton of liquid scintillator (LS) contained in an acrylic vessel 35.4~m in diameter. The structure is substained by a stainless steel truss 40.1~m in diameter. The scintillation light following an energy deposition in the detector will be read by 17,612 20-inch photomultiplier tubes (PMTs) and 25,600 3-inch PMTs. The detector, the PMTs, and the connected electronics will be immersed in a water pool, 41~m in diameter and 41.5~m in height, which will serve as a water Cherenkov detector with additional 2,400 20-inch PMTs. A detailed description of the experimental facility can be found in~\cite{JUNO:2021vlw} and references therein.
The main goal of JUNO is the determination of the neutrino mass ordering through the study of the oscillated antineutrino flux coming from the Yangjiang and the Taishan nuclear power plants, both located at the optimized distance of about 53~km. Given its dimensions and anticipated performance, JUNO has a vast and unique physics program~\cite{JUNO:2021vlw}.

\begin{figure}[h]
\centering
\includegraphics[width=\textwidth]{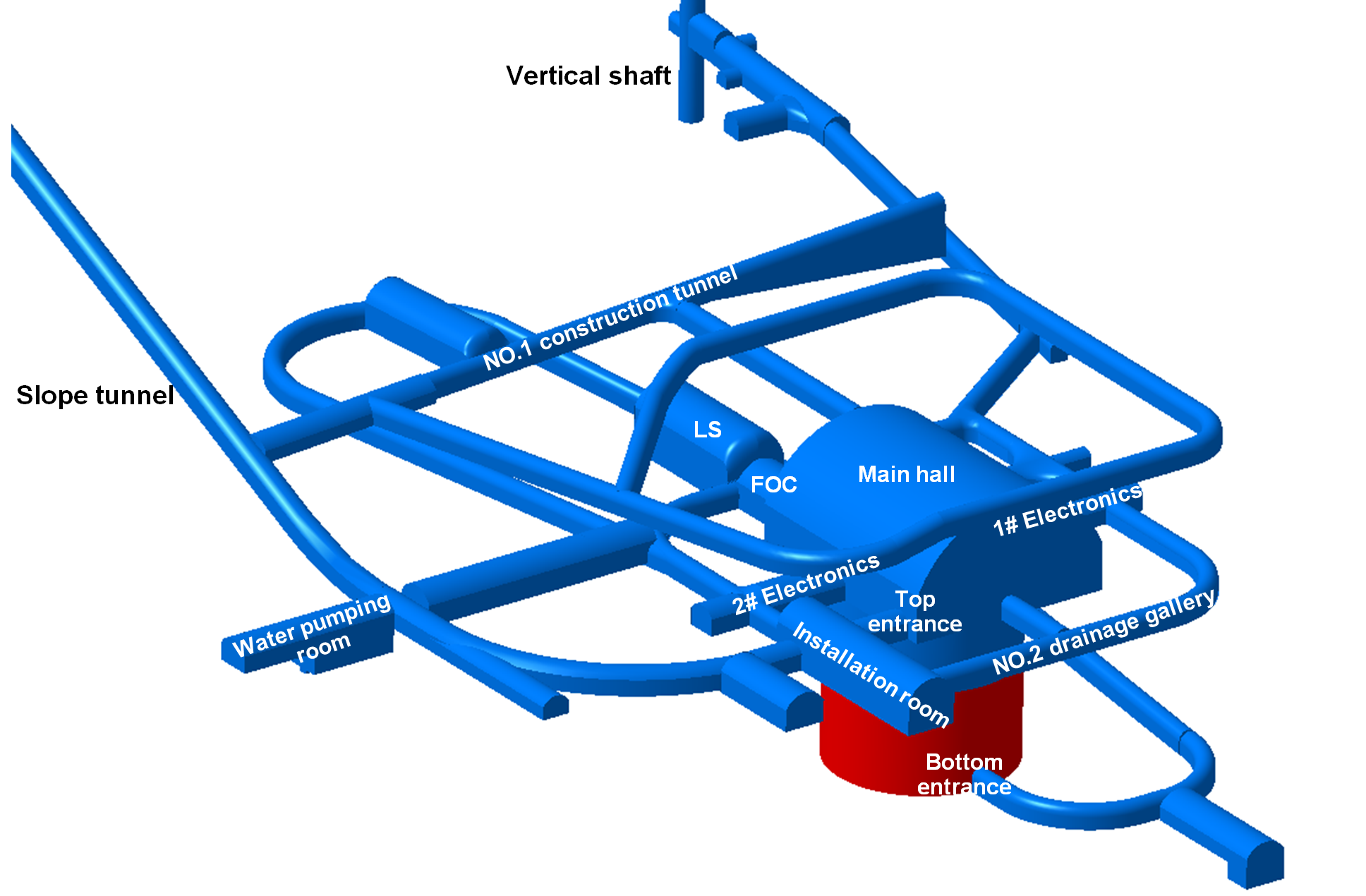}
\caption{The layout of the JUNO underground laboratory.} \label{fig:layout}
\end{figure}

The underground laboratory hosting JUNO is located in Jinji town, 43~km southwest of the city of Kaiping, in Guangdong province, China. The geographic location is 112$^{\circ}$31'05'' E and 22$^{\circ}$07'05'' N. The civil constructions finished at the end of 2021, and the total underground volume is about 300,000~m$^3$. The underground facility has two accesses: a 564~m deep vertical shaft and a 1266~m long tunnel with a slope of 42.5\% (see Fig.~\ref{fig:layout}). The surrounding rock is granite with a measured average density of 2.61~g/cm$^3$. The detector is located in a cylindrical pit inside the main experimental hall, which has a size of 45.6~m\,$\times$\,45.6~m\,$\times$\,71.9~m (height) with an arched top, for a total volume of about 120,000~m$^3$. The vertical overburden at the center of the JUNO detector is equal to 693.35~m ($\sim$1800~m.w.e.). There are two entrances to the main hall: one provides access to the area around the top of the pit, while the other is a gateway to the bottom of the water pool, needed during the construction phase. Both accesses deliver an air shower to people and goods entering the detector area. There are several rooms around the top level of the main experimental hall, some of which are directly connected to it and can be accessed both from inside the hall or through the respective entry doors facing the various connection tunnels (see Fig.~\ref{fig:layout} and Fig.~\ref{fig:ventilation}). The room hosting all pipes and tanks needed for the the LS filling, circulation and overflow control (FOC room), the two rooms with the electronics modules (1\# Electronics and 2\# Electronics), and the two rooms with the fresh air cabinets (1\# Cabinet and 2\# Cabinet) are all connected to the main hall and can be accessed from the top level of the experimental hall. The room for the LS purification (LS room) and a refuge room, instead, can only be accessed through the connection tunnels. A 3-D layout of the underground facility can be seen in Fig.~\ref{fig:layout}, while Fig.~\ref{fig:ventilation} is a map of the top level of the underground laboratory.
By design, the fresh air for the whole underground facility will be sucked in from above ground, through dedicated cabinets at the top of the vertical shaft, and then exhausted through the slope tunnel. However, the above ground infrastructure is still under construction and expected to be ready by summer 2023. Since the installation of the JUNO detector started at the beginning of 2022, during the first months of activities in the underground laboratory the fresh air was temporarily drawn from the bottom of the vertical shaft by means of powerful fans. For the ventilation within the main experimental hall, there are two fresh air inlet cabinets on the upper level, while air is exhausted through dedicated orifices at the bottom of the hall.

The environmental radon concentration in the main hall during the JUNO
detector construction should be kept around 100~Bq/m$^3$ to minimize radon daughter depositions on the detector. The total volume of JUNO underground laboratory is larger than that of LNGS and a little smaller than that of SNOLAB. However, the main hall at JUNO is the biggest one in the world. The upper limit of ventilation power for fresh air in the JUNO experimental hall is about 40,000~m$^3$/h due to civil construction, which is only a third of SNOLAB. Therefore, the achievement of a low  radon environment through ventilation is more challenging at JUNO. At the start of the underground operations, the radon concentration in the experimental hall reached 1600~Bq/m$^3$. In order to effectively improve the situation, the underground radon sources were carefully investigated.

The paper is organized as follows: the study of the underground radon sources by means of benchmark experiments in the refuge room are described in Sec~\ref{sec2}. Optimization of the ventilation system in the underground tunnel and in the main hall are discussed in Sec~\ref{sec3}. The decomposition of the radon contributions to the main hall under steady state conditions are reviewed in Sec~\ref{sec4}. Finally, the application of the JUNO experience to other experiments is summarized in Sec~\ref{sec5}.

\section{Study of underground radon sources in the refuge room}
\label{sec2}

\begin{figure}[h]
\begin{center}
	\includegraphics[width=0.5\textwidth]{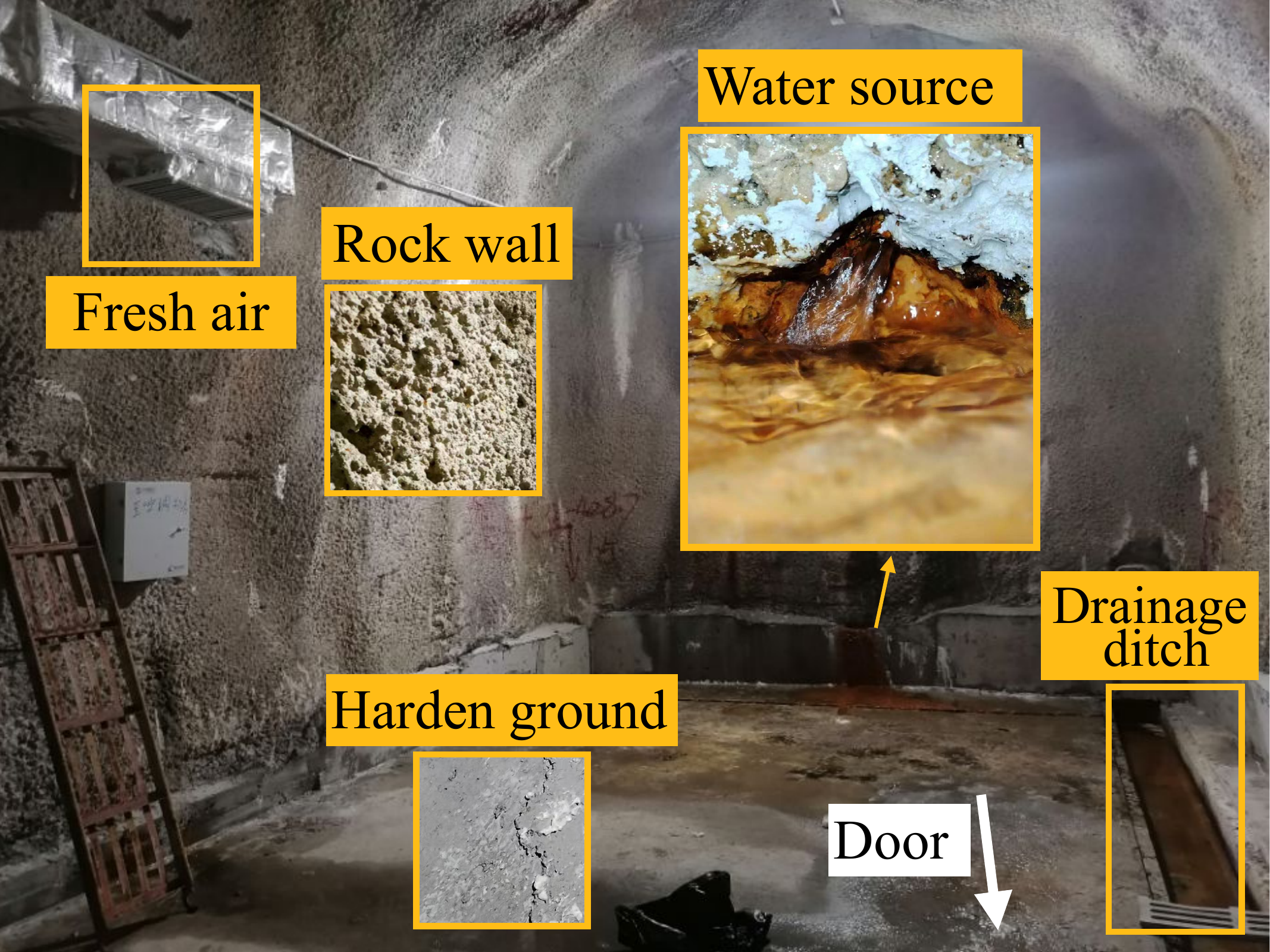}
    \caption{The configuration of the refuge room.}\label{fig:refugeRoom}
\end{center}
\end{figure}

The $^{238}$U concentration in the rock at the JUNO site was measured to be about 120~Bq/kg~\cite{JUNO:2021vlw}, which is 1-2 orders larger than SNOLAB~\cite{Strati:2017rkj} and LNGS~\cite{Coltorti:2011gr}, and $^{222}$Rn, as its daughter, can emanate from the rock surface. Moreover, there is a large amount of underground water at the JUNO site, and the displacement can reach 450~m$^3$/h. 
A large amount of radon from $^{226}$Ra decay is accumulated in the water. The partition factor of radon between water and air is about 0.25 at 20$^{\circ}$C, so radon in water can also diffuse to the air. To better understand the main radon sources in the underground air, we have done some tests in the refuge room (4.5~m\,$\times$\,9~m\,$\times$\,4.7~m (height) with an arched top) near the main hall. As shown in Fig.~\ref{fig:refugeRoom}, in the refuge room there are rock walls, harden ground, water outflow, drainage ditch, and fresh air inlet, in a configuration quite similar to the one of the main hall.

The radon content of the air in the refuge room is constantly fed by the emanations of rock and water, while the inlet of fresh air with low radon level by the ventilation system can dilute it. Thus the radon emanation rate can be expressed in kBq/h as:
\begin{equation}
E = E_{\rm rock}+E_{\rm water}+\Phi C_{\rm fresh}, \\
\label{equ:ema}
\end{equation}
where $E_{\rm rock}$ and $E_{\rm water}$ are the radon emanation rates from the rock and the water outflow, respectively. $\Phi$ is the rate of fresh air inlet in m$^3$/h, and $C_{\rm fresh}$ is the radon concentration in the fresh air.

Due to the imperfect sealing of the refuge room, radon diffuses from inside to outside, and vice versa. The net diffusion depends on the difference in radon concentration on the two sides. On the other hand, radon decay decreases its concentration with a half-life of 3.8~days. The differential function of the radon concentration in the air, $C_{\rm air}$, can be described as:

\begin{equation}
\frac{dC_{\rm air}}{dt} = \frac{E}{V}-(\frac{L+\Phi}{V}+\lambda)C_{\rm air}\\
\label{equ:diff}
\end{equation}
where $L$ is the radon net diffusion rate in m$^3$/h, $V$ = 175~m$^3$ is the volume of the refuge room, $\lambda$ = $7.6\times10^{-3}$~h$^{-1}$ is the decay constant of $^{222}$Rn. The solution of Equ.~\ref{equ:diff} is obtained as

\begin{equation}
C_{\rm air}(t) = \frac{E(1-e^{-\lambda_et})}{V\lambda_e}+C_0e^{-\lambda_et}, \lambda_e = \frac{L+\Phi}{V}+\lambda. \\
\label{equ:RnCon}
\end{equation}
where $C_0$ is the initial concentration of radon.
The radon concentration in air at equilibrium is obtained as:
\begin{equation}
C_{\rm air}(\infty)=\frac{E}{L+\Phi+V\lambda}.
\label{equ:equilibrium}
\end{equation}

The radon emanation from the rock of the refuge room was measured with the RAD7 instrument~\footnote{Details about this measurement are discussed in the appendix.}~\cite{RAD7}: it is 3.6$\pm$0.5~Bq/m$^2$/h for the 114~m$^2$ wall surface, while it is 1.9$\pm$0.5~Bq/m$^2$/h for the 60~m$^2$ concrete harden ground. Therefore, the total radon emanation rate from the rock, $E_{rock}$, is 0.52$\pm$0.06~kBq/h, which would lead to about 400~Bq/m$^3$ radon concentration in the air of the refuge room at steady state equilibrium, as calculated by Equ.~\ref{equ:equilibrium} assuming no ventilation and no air leakage ($\Phi=L=0$ in Equ.~\ref{equ:equilibrium}).

The measurement of radon in water can be realized with the accessory RAD AQUA~\cite{RAD7}. The sucked in water is sprayed by AQUA, thus increasing the efficiency of radon emanation from water; then, the radon gas is pumped to RAD7 for the measurement. To obtain a reliable result, several hours are required to reach the equilibrium condition. The radon concentration in water can be calculated from the radon partition factor $a$ between water and gas, which is a function of temperature $T$~\cite{RAD7}:

\begin{equation}
a = 0.105+0.405\times e^{-0.0502T} \\
\label{equ:factor}
\end{equation}

We measured that the temperature difference between RAD AQUA and RAD7 is about 1$^{\circ}$C, which leads to a fluctuation of about 2.3\% in the partition factor $a$. Conservatively, we take 3\% as the systematics for the measurement of the radon concentration in water due to the temperature uncertainty. The radon content in the water exiting the rock was measured to be 75.0$\pm$0.7~kBq/m$^3$, while the one in the drainage ditch water was 22.3$\pm$0.6~kBq/m$^3$. A large amount of radon from water emanated into the air quickly after the outflow.

The initial radon concentration $C_0$ was measured as $36.0\pm6.6~{\rm Bq}/\rm m^3$ with a ventilation of 600~m$^3$/h fresh air. After that, a few actions were taken with the radon concentration in air recorded every three hours with the RAD7.

\begin{itemize}

\item The fresh air was turned off (thus $\Phi=0$ in Equ.~\ref{equ:RnCon} and Equ.~\ref{equ:equilibrium}) and all the known exits of the refuge room, including spaces around the door, a drainage ditch and fresh air pipes, were blocked. The measured radon concentration in the refuge room air as a function of time is shown in Fig.~\ref{fig:refuge1}~(a) and it reached the equilibrium value of about 2700~Bq/m$^3$ after $\sim$100~hours. The model described in Equ.~\ref{equ:RnCon} was used to fit the data points, with radon emanation from water outflow and radon net diffusion rate as free parameters. The uncertainties on the measurements of radon emanations from rock and water are used as the systematics of this fit. The best fits are $E_{\rm water}$ = 16.6$\pm$0.3~kBq/h and $L$ = 4.8$\pm$0.1~m$^3$/h including the systematics. This shows that the radon emanated from the water outflow is about 30 times larger than the radon emanated from the rock. The decreasing rate of the radon concentration in air due to radon diffusion out of the refuge room is about four times larger than the decay constant, shortening the time to reach equilibrium by four times.

\begin{figure}[h]
\centering
  \subfloat[]{\includegraphics[width=0.45\textwidth]{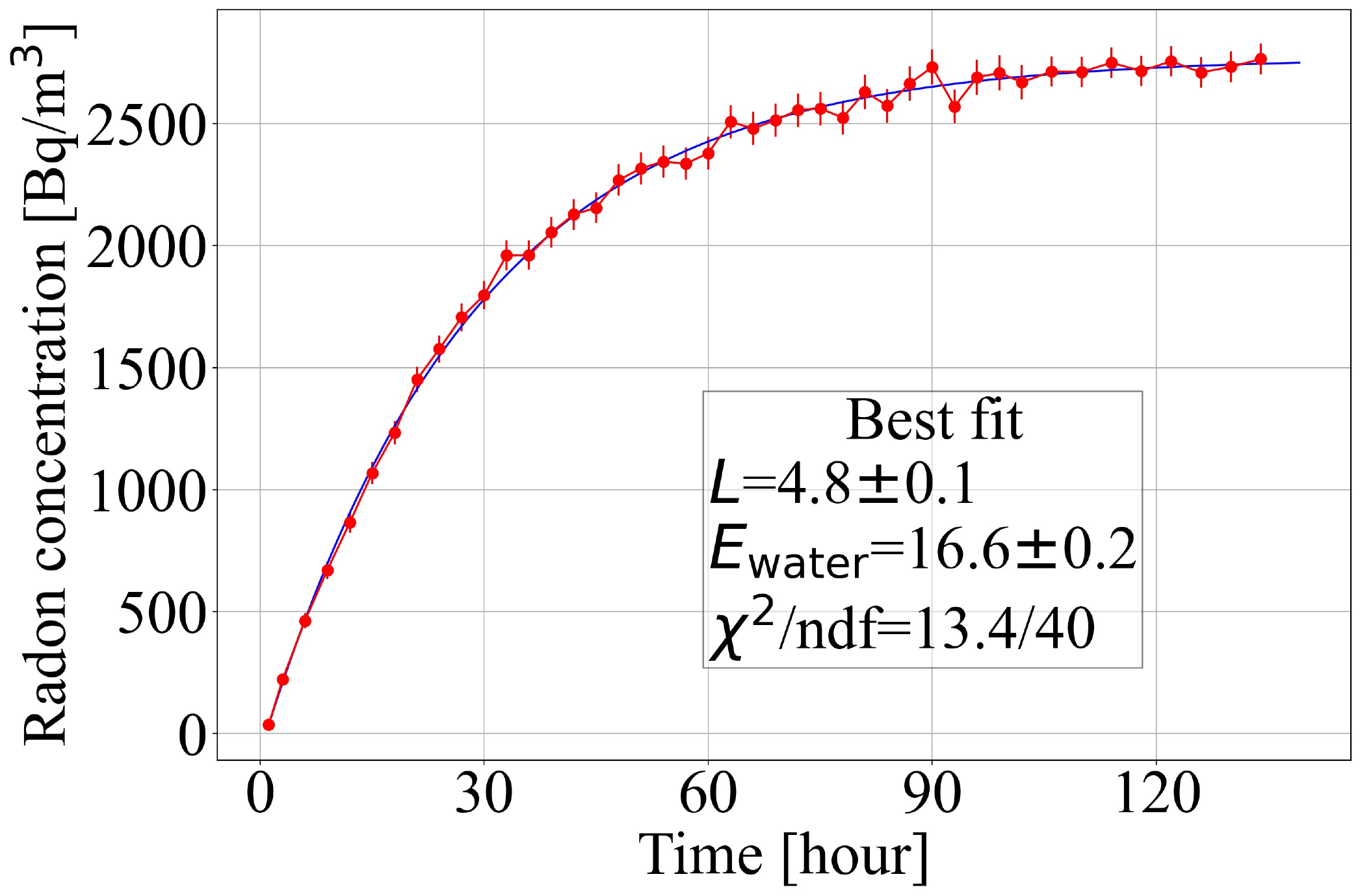}}
  \hfill
  \subfloat[]{\includegraphics[width=0.45\textwidth]{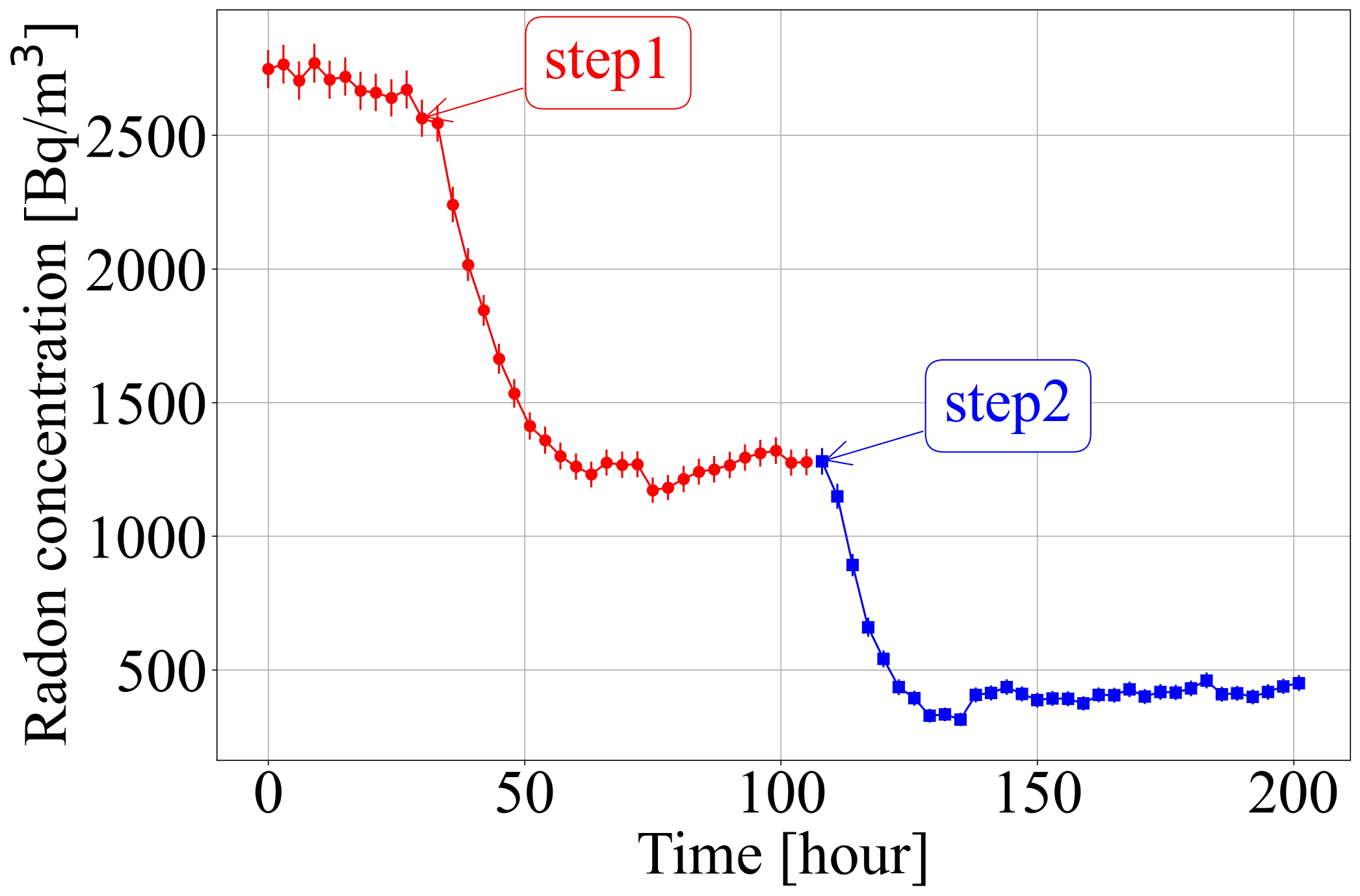}}
\caption{(a) The radon concentration in the refuge room air has reached the equilibrium state of about 2700~Bq/m$^3$ after $\sim$100~hours. The model described in Equ.~\ref{equ:RnCon} is used to fit the data points, while radon emanation from water outflow and the radon diffusion rate are two free parameters. The uncertainties for the best fits are only statistical. (b) The radon concentration after opening the crack on the bottom of the refuge room door (step 1) is shown as the red circle curve, and the radon concentration after exhausting the water outflow from one point (step 2) is shown as the blue triangle curve. }\label{fig:refuge1}
\end{figure}


\item The blocking of the gap (about 0.02~m$^2$) at the bottom of the refuge room door was removed to quantify the relationship between the radon diffusion rate and the dimension of the leak area. The radon concentration in air decreased to a new equilibrium value of 1258$\pm$45~Bq/m$^3$ after 30 hours, shown in the red curve of Fig.~\ref{fig:refuge1}~(b). Since the total radon emanation from rock and water outflow was not changed, the radon diffusion rate was calculated as 12.3$\pm$0.6~m$^3$/h according to Equ.~\ref{equ:equilibrium}.

\item One of the most significant water outflow was exhausted outside of the refuge room by a flexible tube through the crack at the bottom of the entrance door. In this way, the contribution from water outflow can be quantified. We measured the water flow from the same outlet at different times recording a fluctuation of about 20\%. The water outflow from this point was measured to be 0.12$\pm$0.03~m$^3$/h, where the uncertainty includes the flowmeter precision and the flow rate fluctuation. The radon concentration in the air of the refuge room further decreased to 403$\pm$35~Bq/m$^3$ after about 20~hours, as shown in the blue curve of Fig.~\ref{fig:refuge1}~(b). The residual radon emanation from water was calculated as 5.0$\pm$0.5~kBq/h according to Equ.~\ref{equ:equilibrium}. Therefore, the radon emanation from this water point amounts to 11.6$\pm$0.6~kBq/h. Considering the radon concentration in the fresh water and the ditch, the radon emanation from this water point can be calculated to be 6.3$\pm$2.0~kBq/h, which is about 46\% lower than the measurement.

\item Rock is a heat source inside the refuge room, and the air convection is quite complicated in the room volume. The radon concentration in different places of the room was measured to study the radon uniformity: we measured areas near the water source (169$\pm$23~Bq/m$^3$), near the ditch (129$\pm$35~Bq/m$^3$), and a drier place in the center of the room (64$\pm$17~Bq/m$^3$). The radon uniformity can be assumed around 60\%. The radon monitoring setup was kept at the center of the room to acquire the data points of Fig.~\ref{fig:refuge1}, so the radon concentration around the RAD7 can be affected by the convection. Therefore, the 46\% difference between the measurement and the calculation of the radon emanation from water is acceptable.

\end{itemize}

To summarize, besides the well-known surface of the rock, radon emanated from the water was also identified as one of the main radon sources in the underground air. It could even play a dominant role, like in this refuge room experiment, depending on the water outflow rate and the radon concentration in the water. In addition, it is challenging to seal a room and thus the radon diffusion can be significant. The radon concentration in the refuge room decreased by a factor of two due to 0.02~m$^2$ leak area, while the diffusion effect plays an opposite role in the main experimental hall, which will be discussed in the next section. Finally, ventilation is the most effective way to reduce radon coming  not only from the rock but also from the water: it is important and complicated to design a good and effective ventilation system in such a large underground laboratory.

\section{Underground ventilation design}
\label{sec3}

\subsection{Ventilation in the tunnel}
\begin{figure}[h]
\begin{center}
	\includegraphics[width=\textwidth]{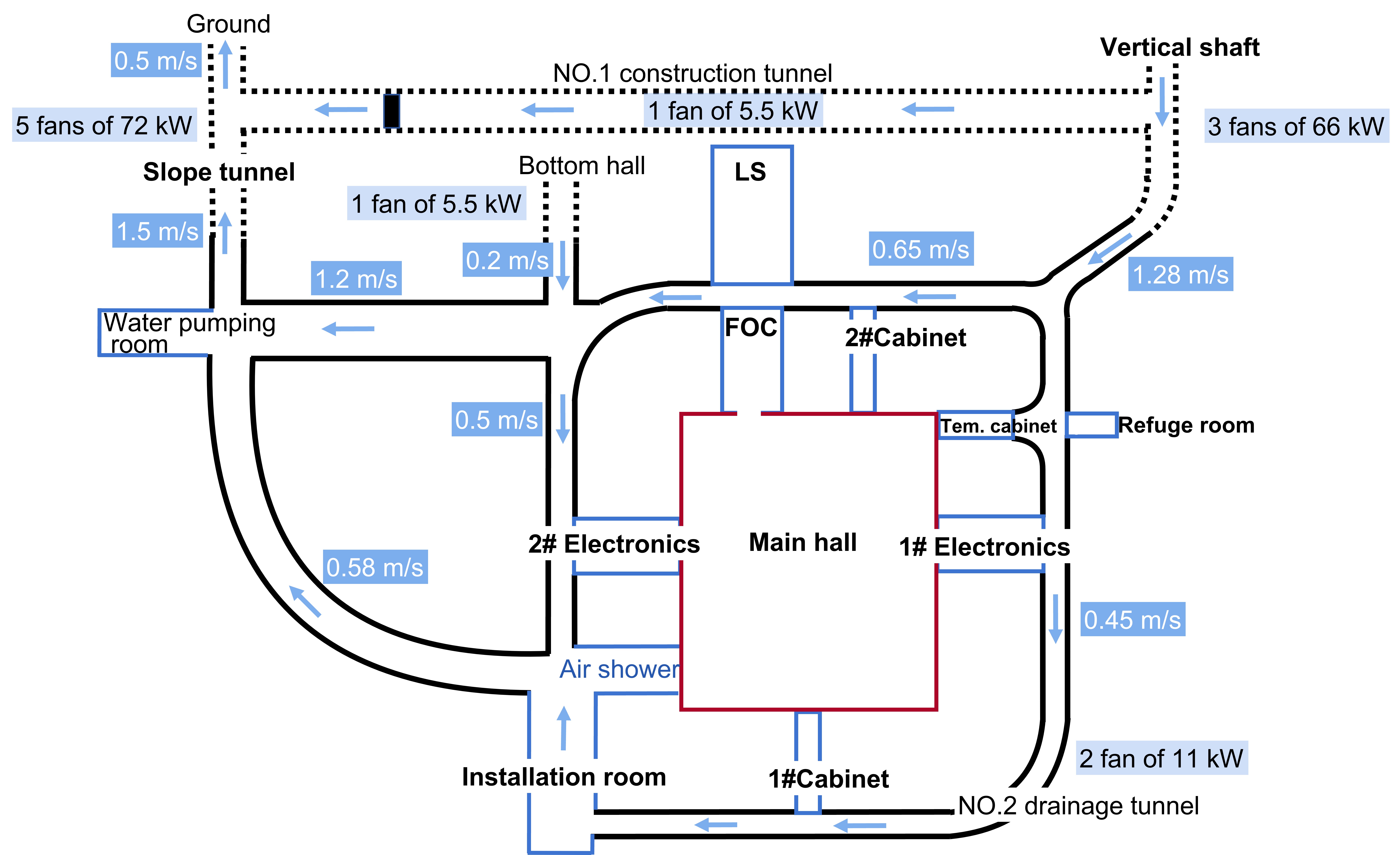}
    \caption{The wind speed in the tunnels after deploying twelve fans with a total power of 156~kW, each one labeled in the figure.}\label{fig:ventilation}
\end{center}
\end{figure}

A good ventilation is mandatory to keep radon concentration under control in the underground air. Twelve fans, for a total power of 156~kW,  were deployed in the tunnels of the JUNO laboratory to improve the ventilation underground: the final layout of the wind direction and speed is shown in Fig.~\ref{fig:ventilation}. Already by this expedient, the radon concentration in the main hall decreased to 200-800~Bq/m$^3$.

\begin{figure}[h]
\begin{center}
	\includegraphics[width=\textwidth]{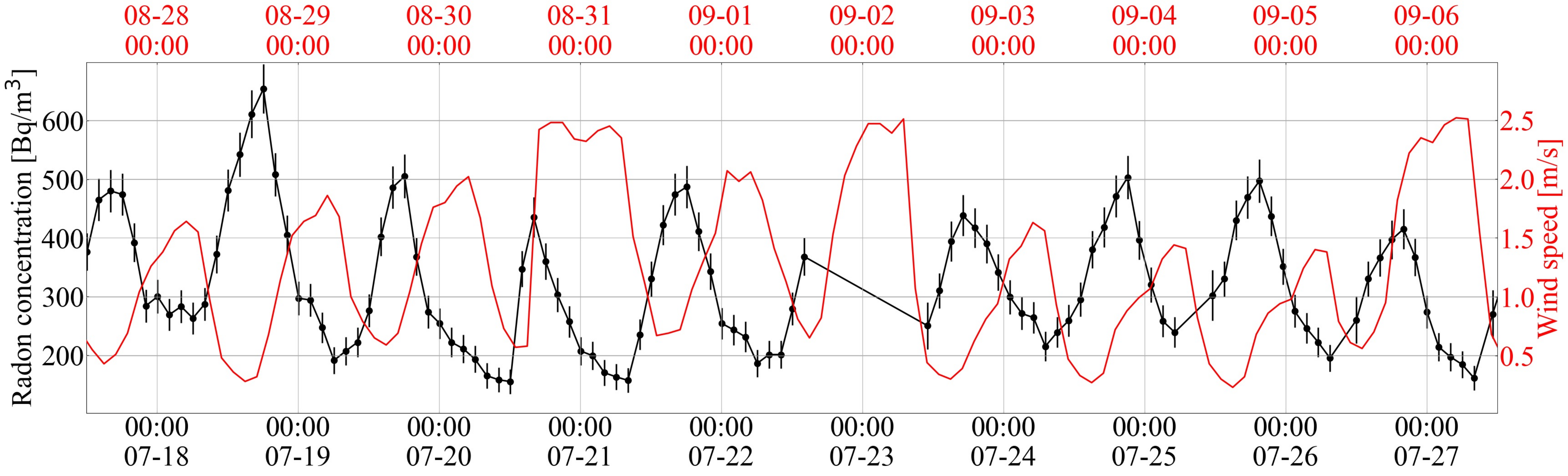}
    \caption{The monitoring of radon concentration in the air of the JUNO underground experimental hall is shown as the black data, and a clear oscillation from day to night is observed. The variation of wind speed underground is shown as the red line. There is a negative correlation between radon concentration in the main hall and the wind speed.}\label{fig:RnWind}
\end{center}
\end{figure}

Moreover, a clear oscillation of radon concentration from day to night was observed, as shown in Fig.~\ref{fig:RnWind}. The radon concentration is higher during the day (with a maximum in the afternoon), while it becomes lower during the night (with a minimum in the early morning). This phenomenon is opposite to that observed above ground, hourly concentrations tended to decrease during the day to a minimum in the late afternoon, and increase thereafter to a maximum concentration in the early
morning~\cite{SESANA2003147}. The explanation for the oscillation observed above ground is that the daily higher temperature leads to a maximum extension of the atmospheric mixing layer, thus lowering the radon concentration at a height of about 2.5~m from the ground.

\begin{figure}[h]
\begin{center}
	\includegraphics[width=\textwidth]{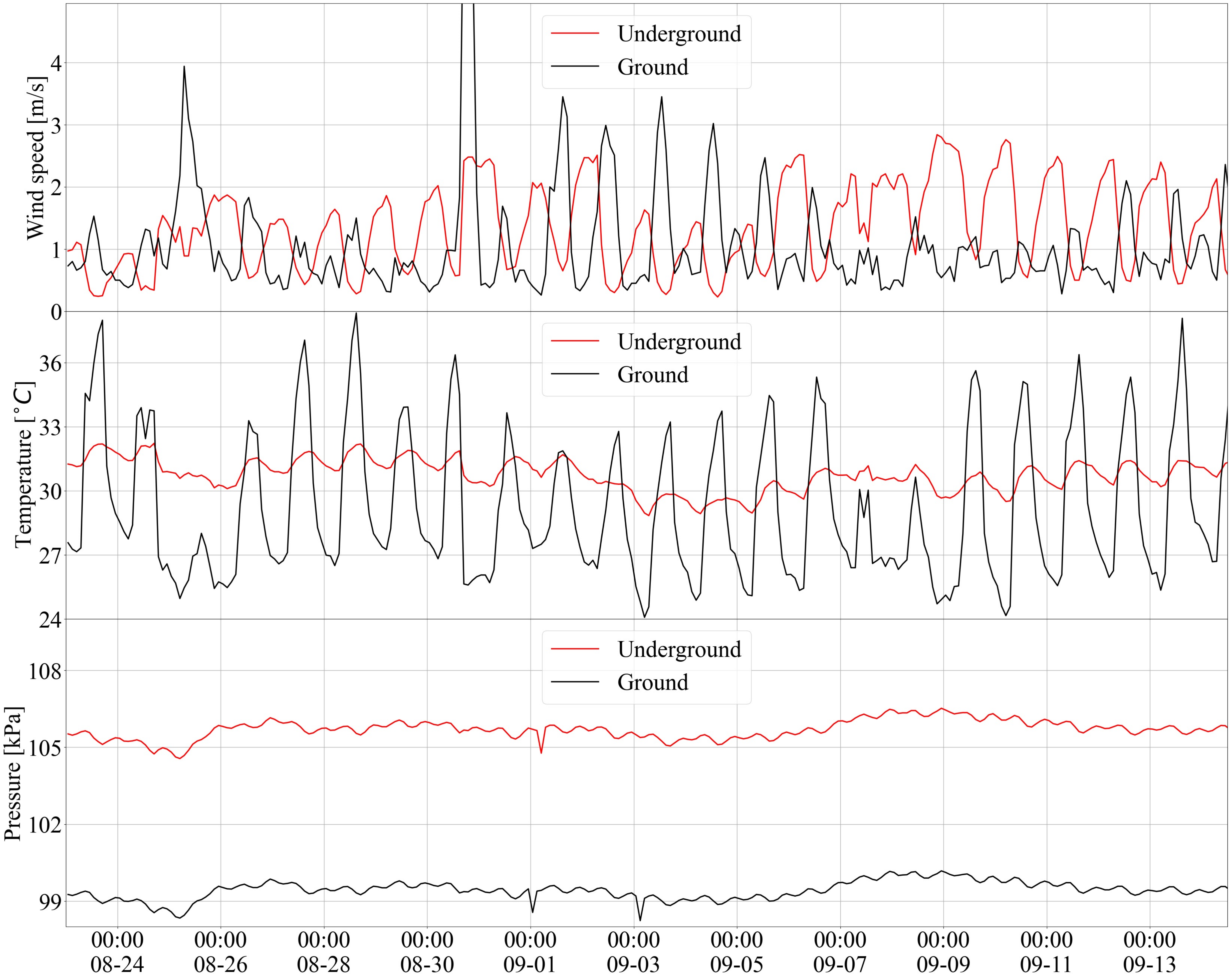}
    \caption{(a) There is a negative correlation between the wind speed above ground and underground. (b) There is a negative correlation between the wind speed underground and the temperature above ground, and the variation of temperature underground is much smaller than that above ground. (c) There is no obvious correlation between the wind speed underground and the pressure.}\label{fig:wind}
\end{center}
\end{figure}

To understand the radon oscillation phenomenon from day to night, two weather monitoring setups at the top and bottom of the vertical shaft were installed. The variation of wind speed, temperature, and pressure on the ground and underground are shown in Fig.~\ref{fig:RnWind} and~\ref{fig:wind}.
The fluctuation of temperature from day to night is larger above ground than underground, and the variation of temperature underground, at a depth of $\sim700$\,m, is about 1-2\,$^{\circ}$C. From Equ.~\ref{equ:factor}, 1-2$^{\circ}$C variation on the temperature will lead to 4\% difference in the partition factor of radon emanation from water. So, the variation of temperature from day to night underground is not the crucial reason for the oscillation of radon concentration underground. There is an anti-correlation between the wind speed above ground and underground. From the weather information, we found that the higher the temperature above ground, the smaller the wind speed underground, thus the higher the radon concentration underground. No obvious relationship with the pressure was found, which is similar to the conclusions in Ref.~\cite{KAMRA2015170}. Since the weather setups were installed in August 2022, the weather information in August is plotted together with the radon oscillation recorded during the month before (July 2022) in Fig.~\ref{fig:RnWind}. The strong anti-correlation between the radon concentration and the wind speed underground is clearly visible. The wind speed underground oscillated between 0.5~m/s and 1.5~m/s from day to night, with a repercussion on the radon exhaust efficiency in the tunnel and, thus, on the radon concentration in the air. The natural ventilation underground is related to the weather conditions above ground which, therefore, represent the critical factor also for the radon concentration in the underground air.

We screened the radon concentration at various positions along the underground laboratory to find the radon source. The radon concentration near the bottom of the vertical shaft was around 30~Bq/m$^3$, so the fresh air is good enough. At about 100~m away from the bottom of the vertical shaft, the radon accumulated in the No.1 construction tunnel that had worse ventilation. It turned out that here the wind direction was slightly from the slope tunnel towards the vertical tunnel: therefore, the air rich in radon of the No.1 construction tunnel was blown to the vertical shaft tunnel and then to the experimental hall. In fact, the 24-hours monitoring of the radon in the No.1 construction tunnel showed the same oscillated radon behaviour from day to night, with values between 600 and 1100~Bq/m$^3$, consistent with what was measured in the main hall. Therefore, We optimized the deployment of ventilation fans inside and outside the No.1 construction tunnel, to force the wind direction from the vertical shaft towards the slope tunnel. After that, the radon concentration in the air of the tunnel area between the vertical shaft and the LS room decreased to below 100~Bq/m$^3$, while the radon concentration inside the main hall reached values in the range 200-400~Bq/m$^3$. The next step to decrease these radon concentration values, was to optimize the ventilation inside the main hall, as discussed in Sec~\ref{sec3.2}.

\subsection{Ventilation in the main hall}
\label{sec3.2}
There are two fresh air cabinets (1$\#$ Cabinet and 2$\#$ Cabinet in Fig.~\ref{fig:ventilation}) on the top of the main hall for the introduction of the fresh air, while there are additional four temporary cabinets with three stages of filters for the air circulation within the hall, to improve the cleanliness level in the detector area. The temperature inside the main hall is kept stable at (21$\pm$1)$^{\circ}$C by cooling water inside the cabinets. As a result, the radon emanations from rock and water inside the main hall should remain almost stable.

By design, the inlet for the two fresh air cabinets is above ground, and the fresh air is transported underground through a long windpipe along the vertical shaft. However, the construction of this infrastructure is expected to be completed by summer 2023: before that date, the fresh air intake is placed underground near the bottom of the vertical shaft, therefore in a non-ideal condition. There is no sucking power at the inlet of the pipe, so the fresh air is drawn into the main hall only by the cabinets. However, the length of the pipe from the intake at the vertical shaft bottom to the cabinets near the main hall is about 200~m, so the power of the cabinets themselves is not effective to draw fresh air. Therefore, a 22~kW fan was added at the inlet of this pipe to increase the  fresh air stream, together with a primary filter to improve the air quality. The final amount of fresh air supplied by the two cabinets is 35,000~m$^3$/h, 10\% of which is directed towards the LS room.

A large amount of water with a flow rate of 70~m$^3$/h  is continuously  discharged from the rock at the bottom of the water pool: the radon emanating from the water is a significant source inside the experimental hall. To improve this situation, we put  one fan outside the bottom of the hall to help the air exhaust from inside, with a flow rate up to 22,000~m$^3$/h. This is useful not only to remove the huge amount of radon emanating from the water, but also to have a better ventilation inside the water pool.

Compared to the total volume of 120,000~m$^3$ of the experimental hall, the net air flow rate of 10,000~m$^3$/h introduced in the hall (inlet 32,000~m$^3$/h, outlet 22,000~m$^3$/h) may not be enough to maintain a positive pressure with respect to the outside, which is important to avoid dirty air leaking into the experimental area. Therefore, we have converted one circulation cabinet (indicated as Tem. Cabinet in Fig.~\ref{fig:ventilation}) to serve as additional provider of 20,000~m$^3$/h fresh air to the hall. Unfortunately, the air intake of this cabinet is in the No.2 drainage gallery, where a large amount of water is discharged from the rock walls with a flow rate of 150~m$^3$/h, which is about one-third of the whole water displacement underground. Therefore, there would be a strong accumulation of radon: with two 5.5~kW fans positioned in this gallery, the ventilation improved  and the radon concentration decreased from 2000~Bq/m$^3$ to 350~Bq/m$^3$ in the air close to the water source. This is helpful also to avoid high radon leaking into the hall. To further improve the fresh air quality from the temporary cabinet, we extended the intake to the tunnel with
low radon fresh air by a stainless steel pipe.

In summary, we have about 52,000~m$^3$/h of fresh air introduced in the main hall and shared with the FOC and the two electronics rooms (which are directly connected to the main hall). Considering the 22,000~m$^3$/h outlet from the bottom of the main hall, there is a residual flow of 30,000~m$^3$/h to keep a positive pressure inside the main hall, and thus avoid the leakage of air rich in radon from the No.2 drainage gallery into the experimental area.

\subsection{Long time monitoring of radon concentration}
The final design of the ventilation system underground is summarized in Table~\ref{tab:ventilation}, while the radon concentrations measured at different locations are listed in Table~\ref{tab:Rndis}. The radon concentration inside the main hall, including adjacent rooms connected to it, is constantly around 100~Bq/m$^3$; it can reach 140-450~Bq/m$^3$ in some areas with a large amount of underground water and bad ventilation outside the experimental hall.

\begin{table}[htbp]
\centering
 \caption{ \label{tab:ventilation} The optimized ventilation inside the experimental hall at the JUNO site from October 2022. The area of the surface includes both the rock and floor, and the ventilation is calculated based on the wind speed shown in Fig.~\ref{fig:ventilation}.}
 	\begin{tabular}{c|c|c|c|c}
        \hline
    & Water outflow [m$^3$/h] & Volume [m$^3$] & Surface [m$^2$] & Ventilation [m$^3$/h]\\ \hline
    Main hall & Top: 0.4, bottom: 70 & 120,000 & 5500 & 52,000 \\ \hline
    LS room & 4 & 2700 & 650 & 3000 \\ \hline
    Tunnel & 450 & 180,000 & - & 100,000 \\
    \hline
    \end{tabular}
\end{table}

\begin{table}[htbp]
\centering
 \caption{ \label{tab:Rndis} The radon concentration at different locations underground were measured in Oct.~2022. The label for each location can be found in Fig.~\ref{fig:ventilation}.}
	\begin{tabular}{c|c||c|c}
        \hline
    Location & $^{222}$Rn in air [Bq/m$^3$] & Location & $^{222}$Rn in air [Bq/m$^3$]\\ \hline
    Main hall & 84$\pm$17 &  1$\#$ Electronics & 67$\pm$44\\
    LS room & 69$\pm$15 & 2$\#$ Electronics & 107$\pm$50 \\
    FOC room & 62$\pm$35 & Installation room & 447$\pm$52 \\
    1$\#$ Cabinet & 135$\pm$57 & Bottom of vertical shaft & 28$\pm$19\\
    2$\#$ Cabinet & 43$\pm$36 & Bottom of slope tunnel & 196$\pm$39\\
    \hline
    \end{tabular}
\end{table}

\begin{figure}[h]
\begin{center}
	\includegraphics[width=\textwidth]{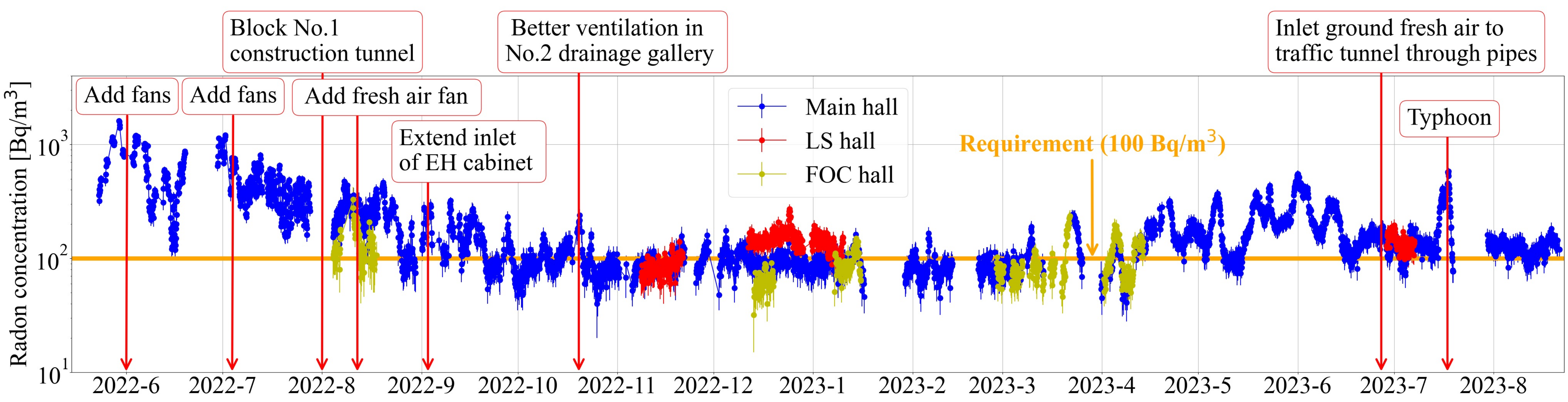}
    \caption{Long time monitoring of radon in the air inside the experiment hall. The main operations are labeled in the figure.}\label{fig:RnMonitor}
\end{center}
\end{figure}

The long-time monitoring of radon in the air inside the main hall is shown in Fig.~\ref{fig:RnMonitor}, and the main operations described in the previous Sections are labeled in the figure. With all of these efforts, the radon inside the experimental hall fluctuated around 100~Bq/m$^3$ constantly since October 2022, which satisfies our requirement. The FOC room is connected to the main hall, so the radon concentration is similar in the two areas. During the first stage of installation works of the  purification plants inside the LS room, the door separating this area from the tunnel was not installed due to the need of a wide access for material transportation, so the radon concentration in the LS room was similar to that in the main tunnel, as shown in Table~\ref{tab:Rndis}. After the installation of the door, a flow of 3000~m$^3$/h of fresh air is constantly supplied since December 2022, and the radon concentration in the LS room reached about 150~Bq/m$^3$, mainly due to the larger ratio of water outflow to the volume of the room compared with the main hall.

\section{Decomposition of the radon sources in the main hall}
\label{sec4}
There are four radon sources inside the experimental hall: radon emanation from rock and water, radon diffusion from outside, and radon from fresh air. The measurements for all the sources inside the main hall are summarized in Table~\ref{tab:RnInEH}, while the radon emanation rate is calculated by Equ.~\ref{equ:ema}.

\begin{table}[htbp]
\centering
 \caption{ \label{tab:RnInEH} Radon measurements inside the main hall.}
 \resizebox{0.8\textwidth}{!}{%
	\begin{tabular}{c|c|c}
        \hline
    \multirow{2}{*}{Rn source} & \multirow{2}{*}{Rn measurement} & Rn emanation rate \\
    &&[kBq/h]\\ \hline
    \multirow{2}{*}{Rock} & normal rock: 4.78$\pm$0.97~Bq/m$^2$/h, 5000~m$^2$ &  24 \\
    & harden ground: 1.88$\pm$0.47~Bq/m$^2$/h, 500 m$^2$ & 0.94 \\ \hline
    \multirow{3}{*}{Water} & fresh water from rock: 120~kBq/m$^3$ & \multirow{2}{*}{42 (top)} \\
    & water in drainage ditch: 20~kBq/m$^3$ & \\
    & water flow rate: 0.4~m$^3$/h (top), 70~m$^3$/h (bottom) & 10$^4$ (bottom)\\ \hline
    Fresh air & \multicolumn{2}{c}{52,000~m$^3$/h with Rn $\sim$30~Bq/m$^3$} \\
    \hline
    \end{tabular}
    }
\end{table}

We have a strong air circulation inside the main hall, about 200,000~m$^3$/h on the top and about 20,000~m$^3$/h in the pool, so the uniformity of the radon concentration in the hall volume is quite good. On the top of the hall, the radon emanation rate from rock and water is 67~kBq/h. Considering the large amount of fresh air (net air inlet 30,000~m$^3$/h) with low radon introduced in the main hall, the radon diffusion from outside to inside can be neglected in the calculation. So, the radon concentration in the air at equilibrium can be calculated by Equ.~\ref{equ:equilibrium} to be $\sim$30~Bq/m$^3$, dominated by the fresh air.

However, the average radon inside main hall is about 100~Bq/m$^3$, so the contribution from the bottom radon source must be non-negligible. The rock side of the pool is hardened with 50~cm of concrete and covered with 5~mm HDPE (high density polyethylene) film, so the radon emanation from the rock is negligible. Unlike the HDPE film on the rock side, the HDPE on the pool ground is not fixed to the ground, and there are some drains on the ground under the HDPE film. Moreover, there is a gap between the HDPE film on the side rock and the ground. During the detector construction phase, the drains are always blocked due to the saliferous underground water, so the drains need regularly cleaning and the film is not fixed to the ground. After the installation of the whole detector, all the film will be welded together. The water flow rate is 70~m$^3$/h at the bottom of the hall, which is more than two orders higher than that on the top. All this water is discharged through pipes on a drainage ditch near the exit door. The radon from the water under the HDPE film can diffuse into the air at the edges and through holes of the film, and then circulate to the top. The addition of a fan outside the bottom door to exhaust 22,000~m$^3$/h air from the main hall, helps moving to the outside most of the radon-rich air close to the door.

Unfortunately, starting from April 2023, the situation became worse due to season transition to summer. The ventilation is affected by the difference on the wind speed between summer (1.5~m/s on average in June) and winter (3.0~m/s on average in February). The weather is always sunny and the wind speed underground is smaller than 1~m/s for most of the increasing spikes in Fig.~\ref{fig:RnMonitor}. The radon concentration always decreased suddenly with a strong rain storm, when the temperature started decreasing, which will affect the wind speed underground. In addition, the cabinet room above ground at the vertical shaft started construction, which will also affect the ventilation. The radon concentration in the installation room, used for pre-assembly of the stainless steel supporting bar and storage of acrylic panels, even increased to 1000~Bq/m$^3$. The installation room is at the exit of the No.2 gallery, and there is a large amount of water flowing; even with optimized configuration of fans, the improvement on radon concentration in the installation room is limited.

The fans above ground at the vertical shaft infrastructure were ready on June 13, 2023. Even though the cabinet was not fully ready, we started providing the fresh air by those fans through pipes installed along the vertical shaft to the tunnel underground. After that, the radon concentration in the tunnel decreased from 200~Bq/m$^3$ to around 50~Bq/m$^3$, while the fresh air for the main hall was still sucked in from the bottom of vertical shaft. The radon concentration in the main hall decreased from 300~Bq/m$^3$ to 100-200~Bq/m$^3$, while it was 300-400~Bq/m$^3$ in the installation room. After two weeks, the pipes from vertical shaft were connected to the pipes providing the fresh air to the main hall, thus the 40,000~m$^3$/h fresh air flow at 24$^{\circ}$C from the above ground cabinet is transported to the underground main hall directly. However, the fresh air for the tunnel is still taken at the bottom of vertical shaft with the help of fans, and this is affected by the weather. To avoid the effect of the unstable tunnel air in the radon concentration in the main hall, we stopped the inlet of fresh air from the tunnel through the Tem. Cabinet, shown in Fig.~\ref{fig:wind}. Now the radon concentration in the main hall is around 100~Bq/m$^3$ for most of the time, except the special period of typhoon. It indicates that such extreme weather can still affect the ventilation and radon concentration in underground air, fortunately, this special period is rare.

\section{Summary}
\label{sec5}
There is high radon (120~kBq/m$^3$) concentration in underground water at the JUNO site, while the $^{238}$U in the rock is about 120~Bq/kg. The water displacement at the JUNO site can reach 450~m$^3$/h. The experiment shows that the radon emanation from water is one of the main sources of radon in underground air.

A good ventilation design is quite important to control the entire radon level underground. It is better to achieve no dead zone, and make sure the wind direction is correct. The wind speed underground is correlated to the wind speed and temperature on the ground, so the ventilation power is different from winter to summer. For the JUNO site, deployment of 156~kW in the tunnel is effective to keep a low radon environment of less than 100~Bq/m$^3$. All the attached rooms around the main hall are blocked by doors, and the ventilation of 40,000-50,000 m$^3$ fresh air can maintain the radon to be around 100~Bq/m$^3$ in the laboratory. For the experiment hall, a perfect drainage system will be much helpful to maintain a low radon environment. If the drainage ditch can be constructed as a closed loop with no contact with the air inside the experiment hall, that will be perfect. If there is some area difficult to a closed loop, it is better to have good ventilation around that area towards outside. The results from this study are common to other underground experiments and mine construction.

\acknowledgments
This work is supported by the Youth Innovation Promotion Association of the Chinese Academy of Sciences, and the Strategic Priority Research Program of the Chinese Academy of Sciences (Grant No. XDA10000000). The authors would like to acknowledge Professor Fr\'{e}d\'{e}ric Perrot from Univ.~Bordeaux and Jose Busto from Univ.~Aix-Marseille for their useful discussions.

\section*{Appendix}
In this appendix, we present the method for the measurement of radon emanation from rock.

The radon emanation from the rock is measured with the RAD7 instrument. A plastic basin is covered on the rock surface, and the edge is sealed with plasticene. There are two holes on the basin, and two tubes are used to connect the basin with the RAD7. In this way, the air inside the basin can be circulated, and the radon concentration can accumulate to reach an equilibrium state. The time needed for the measurement is about two hours, labeled as $T$. The relationship of radon concentration
between neighbouring data points can be calculated based on Equ.~\ref{equ:RnCon} as:

\begin{equation}
C_n = \frac{E(1-e^{-\lambda_eT})}{V\lambda_e}+C_{n-1}e^{-\lambda_eT}\\
\label{equ:rockRn}
\end{equation}
By defining $a=\frac{E(1-e^{-\lambda_eT})}{V\lambda_e}$ and $b=e^{-\lambda_eT}$, both $a$ and $b$ are constant. So the radon concentration between neighbouring data points follows a linear function of $C_n=a+bC_{n-1}$.

\begin{figure}[h]
\begin{center}
	\includegraphics[width=0.6\textwidth]{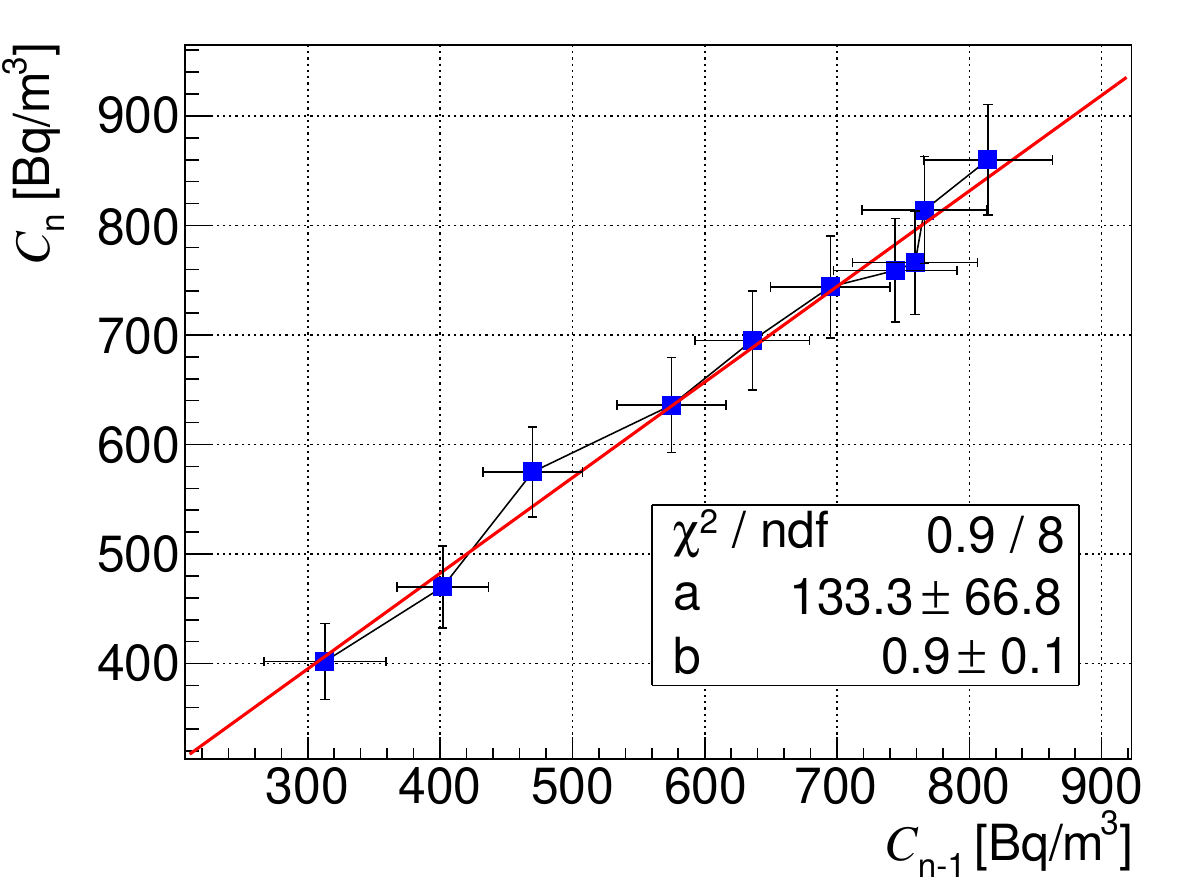}
    \caption{The relationship of radon concentration between neighbouring data points are shown in this figure. A linear curve is used for fitting the data points.}\label{fig:fit}
\end{center}
\end{figure}

The distribution of radon concentration at time t-T and t are shown in Fig.~\ref{fig:fit}, and a linear curve is used to fit the data points. The best fit gives for the radon concentration at time zero the value of 133.3$\pm$66.8 Bq/m$^3$, with a slope of 0.9$\pm$0.1.

The radon emanation rate $E$ can be calculated as $\frac{-aVln(b)}{T(1-b)}$, where $V$ = 2.2 L is the volume of the basin. Based on the best fit results from Fig.~\ref{fig:fit} and on Equ.~\ref{equ:rockRn}, the radon emanation rate from this rock point is calculated to be 3.8$\pm$0.5 Bq/m$^2$/h.

\bibliography{reference}

\end{document}